\documentclass[%
twocolumn,
 superscriptaddress,
 amsmath,amssymb,
 aps, physrev,
]{revtex4-2}

\usepackage{graphicx}
\usepackage{dcolumn}
\usepackage{bm}

\usepackage{siunitx}
\usepackage{upgreek}

\begin{document}

\title{\textbf{Pearl-Vortex Tunneling in Magic-Angle Twisted Graphene} 
}%

\author{Marta Perego}	
\email{mperego@phys.ethz.ch}
\author{Peter Koopmann}
\author{Clara Galante Agero}
\author{Alexandra Mestre Tor\`a}
\author{Artem O. Denisov}
\affiliation{Laboratory for Solid State Physics, ETH Zurich,~CH-8093~Zurich, Switzerland}
\author{Takashi Taniguchi}
\affiliation{Research Center for Materials Nanoarchitectonics, National Institute for Materials Science,  1-1 Namiki, Tsukuba 305-0044, Japan}
\author{Kenji Watanabe}
\affiliation{Research Center for Electronic and Optical Materials, National Institute for Materials Science, 1-1 Namiki, Tsukuba 305-0044, Japan}
\author{Vadim Geshkenbein}
\affiliation{Institute for Theoretical Physics, ETH Zurich,~CH-8093~Zurich, Switzerland}
\author{Gianni Blatter}
\affiliation{Institute for Theoretical Physics, ETH Zurich,~CH-8093~Zurich, Switzerland}
\affiliation{Quantum Center, ETH Zurich,~CH-8093 Zurich, Switzerland}
\author{Thomas Ihn}
\author{Klaus Ensslin}
\affiliation{Laboratory for Solid State Physics, ETH Zurich,~CH-8093~Zurich, Switzerland}
\affiliation{Quantum Center, ETH Zurich,~CH-8093 Zurich, Switzerland}

\date{\today}

\begin{abstract}
Twisted graphene provides a tunable platform for studying superconductivity in
two dimensions. In the presence of electric currents and magnetic fields,
vortices determine the phenomenological properties of the material.  Related
studies usually address bulk properties averaging over ensembles of vortices.
Here, we employ a gate-defined Josephson junction as a single-vortex sensor,
enabling direct access to individual vortex dynamical events.  Our
measurements reveal that, at elevated temperatures ($T>\SI{100}{mK}$),
vortices enter the superconducting leads via classical thermal activation over
energy barriers. At lower temperatures ($T<\SI{90}{mK}$), we observe
macroscopic quantum tunneling through these barriers.  The data are consistent
with a sharp, first-order type quantum-to-classical transition. From our
measurements, we extract vortex entry and exit energy barriers on the order of
a few Kelvin and estimate the barrier thickness to be approximately
$\SI{100}{nm}$, corresponding to about one tenth of the device width.
\end{abstract}

\maketitle


Twisted graphene has recently emerged as a novel superconductor with
remarkable tunability: it can be driven into and out of the superconducting
state only by changing its carrier density \cite{cao2018correlated, lu2019superconductors, cao2018unconventional,
park2021tunable, hao2021electric,
park2022robust, park2022robust, zhang2022promotion,
burg2022emergence}. This property enables the
realization of superconducting nano-devices within a single material, where
adjacent regions can be independently tuned into superconducting, insulating,
or metallic phases \cite{Rodan-Legrain2021, deVries2021, diez2023symmetry,
portoles2022tunable, Ronen2025, wakiri2024tunable,
zheng2024gate}. As a result, twisted graphene provides unprecedented
flexibility for device design and in-situ control of electronic phases.

While much of the existing research has focused either on the microscopic
origin of superconductivity in twisted graphene \cite{Kim2024, Oliver2024,
portoles2024quasiparticle, wu2019phonon, oh2021evidence} or on its
applications in superconducting electronics \cite{deVries2021,
Rodan-Legrain2021, portoles2022tunable, diez2023symmetry, rothstein2025gate,
zheng2024gate, wakiri2024tunable,jha2025large, diez2025probing,Ronen2025}, the
system also constitutes a powerful new platform to study the phenomenology of
superconductivity in two-dimensional (2D) systems. Two-dimensional
superconductors exhibit a number of distinctive features, including quasi-long
range order \cite{Rice1965, MerminWagner1966, Hohenberg1967, Berezinskii1971},
the Berezinskii-Kosterlitz-Thouless transition \cite{Berezinskii1971, KT1973},
and pronounced fluctuation dynamics, here, in the form of topological
excitation, i.e., Pearl-vortices \cite{Abrikosov1957, Pearl1964}. The importance of thermal and quantum fluctuations can be quantified through
the superfluid density $\rho_s$ \cite{Kim2024, Oliver2024} and the sheet
resistance $R^{\scriptscriptstyle \square}$. The low carrier density and large
effective mass associated with flat electronic bands result in a small
$\rho_s$ and a large $R^{\scriptscriptstyle \square}$, which enhance
fluctuation effects in both the thermal and quantum regimes. These properties
make graphene particularly well suited to study vortex dynamics.

\begin{figure*} \includegraphics[width=1.5\columnwidth]{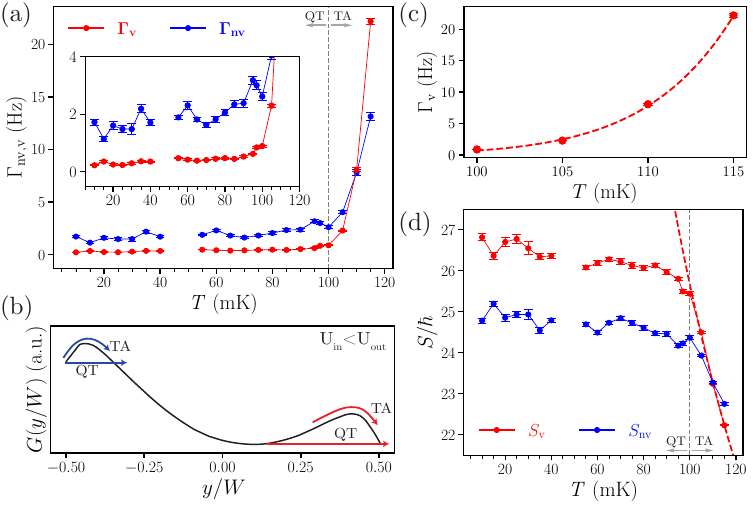} \caption{
(a) Temperature evolution of rates $\Gamma(T)$ for vortex-entry into- and
vortex-exit out of the superconducting leads. (b) Illustration of the
free-energy landscape $G(y/W)$ for vortex motion across the leads. (c) Fitting
the sharp rise in $\Gamma(T)$ at $T > \SI{100}{mK}$ with an Arrhenius law, see
red dashed line, provides a thermal barrier of height $U/k_{\rm
\scriptscriptstyle B} \approx \SI{2.6}{K}$.  (d) The rates saturate at low
temperatures to provide a dimensionless action $S/\hbar = \ln (\nu_0'/\Gamma)$
between 24 and 27. The red dashed line extrapolates the thermal activation law
$S/\hbar = U/k_{\rm \scriptscriptstyle B}T$.
}
\label{Fig1}
\end{figure*}

In this work, we focus on dissipative vortex dynamics at low temperatures,
where macroscopic quantum fluctuations can directly affect the performance of
superconducting devices, including the persistence of currents and the
coherent operation of superconducting qubits.  Previous studies of macroscopic
quantum effects in superconducting junctions have addressed the phase dynamics
intrinsic to the Josephson junction itself \cite{Devoret1985MQT}. Here, by
contrast, we study the macroscopic quantum dynamics of individual
Pearl-vortices \cite{Pearl1964} entering or exiting the superconducting leads.
In our setup, the (gate-defined) Josephson junction then acts as a
single-vortex sensor enabling the detection of the dynamics of
individual Pearl-vortices.

As summarized in Fig.~\ref{Fig1}, we measure the temperature dependence of the
vortex entry and exit rates $\Gamma(T)$, see Fig.~\ref{Fig1}(a), which are
governed by Bean-Livingston edge barriers at the film boundaries
\cite{BeanLivingston1964} (Fig.~\ref{Fig1}(b)).  At elevated temperatures ($T
> \SI{100}{mK}$), the rates increase sharply and are well described by
classical thermal activation over an energy barrier $U$, with $\Gamma = \nu_0
\exp(-U/k_{\rm \scriptscriptstyle B} T)$.  From this analysis, we extract
typical barrier heights $U/k_{\rm\scriptscriptstyle B} \approx \SI{2.6}{K}$
and an attempt frequency $\nu_0 \approx \SI{2.0e11}{Hz}$, see the fit in
Fig.~\ref{Fig1}(c). Upon lowering the temperature, the rates saturate at a finite value, with $\ln
(\nu_0/\Gamma) \approx 27$, deviating substantially from the extrapolated
thermal behavior (Fig.~\ref{Fig1}(d)). We attribute this saturation to the
crossover from thermal activation (TA) to macroscopic quantum tunneling (QT)
of Pearl-vortices, in which the thermal exponent $U/k_{\rm\scriptscriptstyle
B}T$ is replaced by the dimensionless quantum action $S/\hbar$. This crossover
is indicative of a sharp, first-order like quantum-to-classical transition,
consistent with theoretical expectations \cite{Affleck1981,
Chudnovsky1992phase}.  In the following sections, we show how these results
arise from the observation of telegraph-type switching signals produced by
individual Pearl-vortices crossing the superconducting leads near the
Josephson junction sensor.

\begin{figure}[t!] \includegraphics[width=\columnwidth]{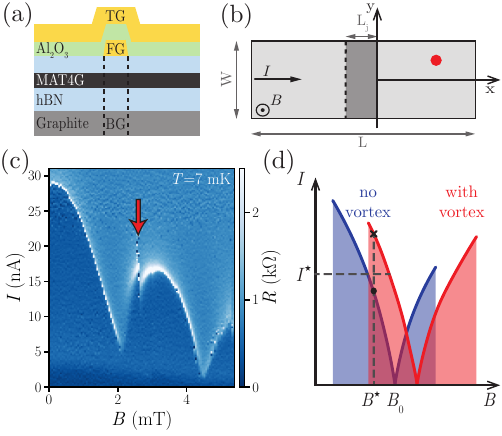}
\caption{
%
(a) Layer structure with gates for device tuning, including a graphite bottom
gate (BG), a gold top gate (TG), and a gold finger gate (FG) defining the JJ.
(b) Device geometry with thickness $d = \SI{1}{nm}$, width $W = \SI{1.1}{\upmu
m}$ along the $y$-direction, and length $L = 6W$ along $x$; the junction width
is $L_{\rm j} = \SI{150}{nm}$.  Vortices penetrate the leads (red dot) and are
detected as jumps in the Fraunhofer pattern.  (c) The field dependence of the
critical current $I_c(B)$ measured at $T = \SI{7}{mK}$ in a film with
`strong-leads' tuning assumes the form of a Fraunhofer-like interference pattern
(FP), see Ref.~\onlinecite{perego2024experimental}. A vortex entering a lead
produces a sudden rightward shift in the Fraunhofer pattern, see red arrow.
(d) Schematic illustrating the change in the dissipation upon vortex entry
into a lead: Measuring the voltage $V$ across the junction at fixed current
$I^\ast$ and field $B^\ast < B_0$, with $B_0$ the zero of the FP, the response
is dissipative when no vortex is present in the lead ($I_c < I^\ast$), while
the junction is superconducting in the presence of a vortex ($I_c > I^\ast$).
The response is reversed when $B^\ast > B_0$.}
\label{Fig2} 
\end{figure}

Our magic angle twisted four-layer graphene (MAT4G) device, with a twist angle $\theta
\approx 1.64^\circ$, consists of an electrostatically defined JJ as
studied previously in Ref.~\onlinecite{perego2024experimental}. A schematic of
the device and its characteristic dimensions are presented in Figs.~\ref{Fig2}, (a) and (b).
All measurements are performed with the carrier density and displacement field in the
superconducting leads set to $n_\textnormal{l} = \SI{4.8e12} {cm^{-2}}$ and
${D_\textnormal{l}/\epsilon_0 = \SI{-0.37}{V/nm}}$, respectively, while the junction region
is held at $n_\textnormal{j} = \SI{6.2e12}{cm^{-2}}$ and
${D_\textnormal{j}/\epsilon_0 = \SI{-0.5}{V/nm}}$ (where $\epsilon_0$ denotes the
vacuum permittivity). Under these conditions, the leads are tuned to the edge
of the superconducting dome, corresponding to the `weak-leads' regime defined in
Ref.~\onlinecite{perego2024experimental}. Owing to the ultra-thin nature of the
device, transverse magnetic screening is negligible, and an applied magnetic field
$H$ penetrates the film nearly uniformly, such that the magnetic induction within
the graphene layers satisfies $B \approx\mu_0 H$ ($\mu_0$ is the vacuum permeability).

When subjecting the device to a perpendicular magnetic field, the maximum
(critical) supercurrent $I_c(B)$ across the junction exhibits a pronounced
field dependence, forming a characteristic Fraunhofer-like interference
pattern (FP) \cite{Tinkham2004,Clem2010}. An example of such a pattern is shown in
Fig.~\ref{Fig2}(c), where the color map is derived from the voltage--current
($V$--$I$) characteristics of the junction. Dark and light blue regions
correspond to low- and high differential resistance $R = dV/dI$, respectively,
distinguishing superconducting ($R=0$) from flux-flow or normal ($R>0$)
regimes.

In our previous study \cite{perego2024experimental}, we observed sharp
discontinuities in the Fraunhofer pattern, which we attributed to the entry or
exit of individual vortices in the superconducting leads, as indicated by the
red arrow in Fig.~\ref{Fig2}(c).  When a vortex enters a lead at $B>0$, it
partially compensates the effect of the applied magnetic field in the vicinity
of the junction, thereby reducing its contribution to the Josephson phase. As
a result, the Fraunhofer pattern shifts toward higher magnetic fields
\cite{Clem2011,KoganMints2014, KoganMints_PC2014}, as schematically
illustrated in Fig.~\ref{Fig2}(d); vortex exit produces the opposite shift. These shifts in magnetic field correspond to changes in flux of order of one
superconducting flux quantum $\Phi_0 = h/2e = 2.07\, 10^{-15}\, {\rm Wb}$ over
an area of order $W^2$. They occur when the vortex manifests sufficiently
close to the junction \cite{KoganMints2014}.  In measurements performed with
the leads tuned to the center of the superconducting dome (the `strong-leads'
regime defined in Ref.~\onlinecite{perego2024experimental}), the energy barriers for
vortex motion are large. This results in well-resolved, individually
observable jumps in the interference pattern and a slow dynamics due to the
formation of metastable states with long lifetimes.

\begin{figure*}[t]
\includegraphics[width=1.5\columnwidth]{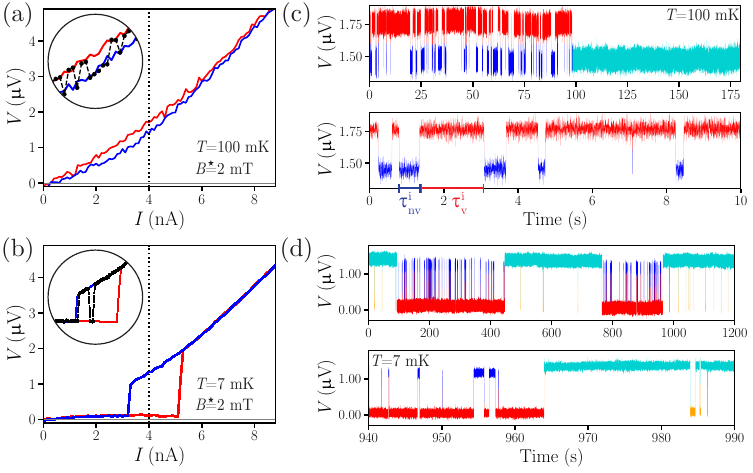}
\caption{
%
(a) $V$--$I$ voltage--current characteristic at $B^\ast = \SI{2}{mT}$ and high
temperature $T = \SI{100}{mK}$ with low- (blue) and high-dissipative (red)
traces. Vortices traversing the leads produce switching between the two
states, see blowup. (b) The same at $T = \SI{7}{mK}$ where the $V$--$I$
characteristic exhibits a sharp critical current $I_c$. (c) Time trace $V(t)$
of the voltage across the junction taken at $B^\ast = \SI{2}{mT}$ and $I^\ast
= \SI{4}{nA}$, see dotted line in (a). The trace $V(t)$ exhibits segments with
switching events (red--blue) due to the passage of vortices across one of the
leads; these `switching segments' are interrupted by `silent' ones (cyan).  (d)
The same at $T = \SI{7}{mK}$ with the previously `silent segments' (cyan in (c)) now exhibiting jumps to the superconducting state (orange).}
\label{Fig3}
\end{figure*}

When tuning the device to the `weak-leads' regime, we observe, in addition to
quasi-stationary shifts due to long-lived trapped vortex states, a distinct
noisy feature in the Fraunhofer interference pattern (see Fig.~4(a) of
Ref.~\cite{perego2024experimental}).  We have attributed this noise to
vortices repeatedly crossing the superconducting leads in the junction
vicinity, which causes jumps in the Fraunhofer pattern. This interpretation is
supported by measurements of the voltage--current ($V$--$I$) characteristics
at a fixed magnetic field $B^\ast = \SI{2}{mT}$ where we observe stochastic
switching between two dissipative states, as shown by the red and blue traces
in Fig.~\ref{Fig3}(a).  At these elevated temperatures ($T = \SI{100}{mK}$),
the superconducting transition is rounded and no sharp critical current can be
identified. Upon cooling to $T = \SI{7}{mK}$ while maintaining $B^\ast =
\SI{2}{mT}$, a well-defined critical current $I_c$ emerges and the switching
occurs between superconducting (red) and dissipative (blue) states, as shown
in Fig.~\ref{Fig3}(b).

We interpret these jumps between distinct $V$--$I$ traces as arising from the
entry and exit of individual Pearl-vortices in the superconducting leads. At
high temperatures, the applied field $B^\ast$ lies to the right of the zero of
the Fraunhofer pattern located at $B_0$ ($B^\ast > B_0$), such that vortex
entry enhances dissipation. At low temperatures, $B^\ast$ falls below $B_0$
($B^\ast < B_0$, see Fig.~\ref{Fig2}(d)) and the entry of a vortex shifts the
junction into the superconducting state producing a downward jump in $V$--$I$. This is
followed by an upward jump upon vortex exit with the junction going back to
the original state.

By measuring the time-dependent voltage $V(t)$ across the junction at fixed
field $B^\ast \approx \SI{2}{mT}$ and current $I^\ast \approx \SI{4}{nA}$, we
observe consecutive voltage jumps consistent with the behavior described
above.  At elevated temperatures $T = \SI{100}{mK}$, see Fig.~\ref{Fig3}(c), we
observe segments exhibiting a telegraph-type switching between high- (red) and
low-dissipative (blue) levels.  These `switching segments', typically lasting
on the order of $\sim\SI{100}{s}$, are interspersed with longer `silent
segments' (cyan) of duration $\sim\SI{1000}{s}$. We interpret this
superstructure in $V(t)$ as arising from an additional quasi-stationary vortex
that enters or exits the superconducting leads on a longer time scale. A
detailed discussion of this behavior will be presented in a forthcoming
publication \cite{Perego2026}.

At low temperatures, the red--blue segments in the voltage trace $V(t)$ are
interrupted by cyan segments that now exhibit telegraph-type switching as
well, visible as orange spikes in Fig.~\ref{Fig3}(d), which shows a
representative section of $V(t)$ measured at $T = \SI{7}{mK}$. In this regime,
the zero of the Fraunhofer pattern has shifted to the right of the measurement
field ($B_0 > B^\ast$), and we therefore associate the superconducting states
(orange and red) with a vortex traversing the lead, while the dissipative
states (cyan and blue) correspond either to vortex-free conditions or to the
presence of a quasi-stationary vortex. 

The interchange of low- and
high-dissipation states associated with the presence or absence of a vortex
appears around a temperature $T \approx \SI{45}{mK}$ where $B_0$ matches the
measurement field $B^\ast$, $B_0\approx B^\ast$. The disappearance of the orange spikes in the cyan--orange segments with
increasing temperature is explained by a decreasing vortex passage time
$\tau_{\rm v}$. Once $\tau_{\rm v}$ drops below the resolution limit of the
detector (with bandwidth $f_{\rm \scriptscriptstyle BW}\approx\SI{1.1}{kHz}$),
the fast dynamics can no longer be resolved, and the cyan segment appears `silent'.

In the statistical analysis of voltage traces with a total duration of 4
hours, we extract the waiting times for vortex entry, $\tau_{\rm nv}$, and
exit, $\tau_{\rm v}$.  Specifically, we define $\tau_{\rm nv}$ as the time
interval without a vortex in the lead and $\tau_{\rm v}$ as the time during
which a vortex is present.  The digitized time traces are analyzed using a
dedicated algorithm \cite{Yuzhelevski_2000}, from which the two-state waiting
times are identified and extracted. These waiting times are subsequently
binned into histograms using the Freedman-Diaconis rule
\cite{Freedman_Diaconis_1981}.  The vortex entry and exit rates, $\Gamma_{\rm
nv}$ and $\Gamma_{\rm v}$, are obtained by fitting exponential distributions. Depending on the type of segment, red--blue or orange--cyan, the pairs of
waiting times $[\tau^i_{\rm nv}, \tau^i_{\rm v}]$ associated with the passage
of the $i$-th vortex typically differ by about one order of magnitude. As a
result, the data are best described by a bi-exponential model, with all rates
corrected for the finite detector bandwidth \cite{Naaman_2006}.  In the
following, we focus on the red--blue segments. 

The temperature dependence of the extracted vortex entry and exit rates
$\Gamma_{\rm nv}$ and $\Gamma_{\rm v}$ is shown in Fig.~\ref{Fig1}(a).
Overall, both rates, $\Gamma_{\rm nv}$ and $\Gamma_{\rm v}$, decrease steeply
with decreasing temperature and saturate at a constant value below
$T\approx\SI{100}{mK}$.  At high temperatures ($T > \SI{120}{mK}$), the
detector performance is limited by thermal noise, while the absence of data
points near $T \approx \SI{45}{mK}$, where $B^\ast \approx B_0$, is due to a
loss of resolution.

We interpret our experimental results within a theoretical framework based on
the free-energy landscape $G(y;H,I)$ \cite{Stejic_1994} that vortices
experience as they are driven across the leads by the applied current $I$, as
illustrated in Fig.~\ref{Fig1}(b).  Vortex penetration into the leads is
hindered by an edge barrier, that is progressively lowered by the applied
magnetic field $H$, while the transport current $I$ introduces a tilt to the
free-energy landscape.  For a homogeneous thin film in the weak screening
limit, where the magnetic induction satisfies $B \approx \mu_0 H$, the free
energy takes the form \footnote{The derivation of Eq.~(1) considers the
situation in a parallel field, that can be adopted to the present situation by
the appropriate replacing of the thickness $d$ with the width $W$.}
\begin{align}
\label{eq: Gibbs}
\begin{split}
   G(y;H,I) & = \varepsilon_0 d
   \left[ \ln\left( \frac{2W}{\pi \xi} \cos\left(\frac{\pi y}{W}\right)\right)
   + 0.38 \right]\\
   & - \varepsilon_0 d\,  \frac{\pi}{2} \frac{\mu_0 H W^2}{\Phi_0}
   \left( 1 - \frac{4y^2}{W^2}\right)-I\frac{\Phi_0 y}{W},
\end{split}
\end{align}
as depicted schematically in Fig.~\ref{Fig1}(b). Here,  $\varepsilon_0 = (4\pi/\mu_0)\,(\Phi_0/4\pi\lambda_{\rm L})^2$ is the vortex
line energy, $\lambda_{\rm \scriptscriptstyle L}$ and $\xi$ are the London
penetration- and coherence lengths.  In our device, the potential landscape
depends on the 2D position $\mathbf{R} = (x,y)$ within the film due to finite
sample size (along $x$) and inhomogeneities, e.g., the superstructure due to
the layer twist or defects and impurities.

At high temperatures, vortices attempting to penetrate (leave) the superconducting leads need to overcome the edge energy barrier
$U_{\textnormal{nv}}$ ($U_{\textnormal{v}}$).
Assuming that {\it{thermal activation}} drives the vortices over the barriers, we
relate the measured rates $\Gamma_{\rm nv,v}$ to the microscopic
parameters $U_{\textnormal{nv,v}}$,
\begin{align} \label{eq:
gamma(U)}
    \Gamma_{\rm nv,v}(T) = \nu_0 \exp \left( -U_{\rm nv,v}/k_{\rm\scriptscriptstyle B} T \right),
\end{align}
with $\nu_0$ the attempt frequency.  Making use of the rates $\Gamma_{\rm
nv,v}(T)$ at high temperatures, we find barriers for vortex entry and exit in
the Kelvin range. For example, analyzing the temperature dependence of the
rate $\Gamma_{\rm v}(T)$ shown in Fig.~\ref{Fig1}(a) in the range
$\SI{100}{mK} < T < \SI{120}{mK}$, we extract $U_{\textnormal{v}}/
k_{\rm\scriptscriptstyle B} \approx \SI{2.6}{K}$ and $\nu_0 \approx
\SI{2.e11}{Hz}$, with the latter value in agreement with previous estimates
\cite{Malozemoff1990, Koshelev1990}.


The rates $\Gamma_{\rm nv,v}$ saturate to constant values upon decreasing the
temperature below $T \approx \SI{100}{mK}$ as seen in Fig.~\ref{Fig1}(a). This
phenomenon is naturally explained in terms of {\it macroscopic quantum
tunneling} \cite{caldeira1983quantum, larkin1984quantum, glazman1984}.  In
this regime, the thermal exponential $U/k_{\rm\scriptscriptstyle B}T$ in
Eq.~\eqref{eq: gamma(U)} has to be replaced by the dimensionless quantum
action $S/\hbar$, such that the rates become
\begin{align}
\label{eq: Gamma(U,S)}
    \Gamma_{\rm nv,v} =\nu_0' 
    \exp \left( -{S_{\rm nv,v}}/{\hbar} \right),
\end{align}
with $S_{\rm nv,v}$ denoting the Euclidean action for vortex entry and exit.
Replotting the data for the rates $\Gamma_\textnormal{nv,n}$ in terms of the
dimensionless action $S_\textnormal{nv,n}/\hbar = \ln (\nu_0'/
\Gamma_\textnormal{nv,n})$ as shown in Fig.~\ref{Fig1}(d), we find nearly
constant values between $24$ -- $27$ at low temperatures $T < \SI{100}{mK}$
(we assume an attempt rate $\nu_0' = \nu_0 \sim \SI{2.0e11}{Hz}$). The values decrease
with increasing $T$ as thermal fluctuations assist the tunneling dynamics
\cite{Affleck1981,larkin1984quantum} at higher temperatures.

The barrier height $U$ for the thermal activation of vortices over the edge
barrier is $U \approx \alpha_{\scriptscriptstyle U} \varepsilon_0 d$, with $\alpha_{\scriptscriptstyle U}$ of order
unity as follows from Eq.~\eqref{eq: Gibbs} and
Ref.~\cite{perego2024experimental}. The vortex dynamics in a superconductor are
dominated by the friction coefficient $\eta_l = \Phi_0^2/2\pi \xi^2 \rho_n$,
where $\rho_n$ is the normal-state resistivity \cite{bardeen1965theory}. The
vortex tunneling action then is given by \cite{larkin1984quantum}
\begin{equation}
   S = \alpha_{\scriptscriptstyle S} \eta_l d q_b^2 = \frac{\alpha_{\scriptscriptstyle S}}{4} \hbar \left(\frac{q_b}{\xi}\right)^2
   \frac{R_{\rm\scriptscriptstyle K}}{R^{\scriptscriptstyle \square}}
\end{equation}
and depends on the barrier width $\sim q_b$ rather than its height $U$.  Here,
$R_{\rm\scriptscriptstyle K} = h/e^2 \approx \SI{25.8}{k\Omega}$ is the
von-Klitzing constant and $R^{\scriptscriptstyle\square} = \rho_n/d$ denotes
the film resistance.  For a metastable potential in the form of a cubic
parabola,
\begin{equation}
   V(q) = 3U\left[\left(\frac{q}{q_b}\right)^2-\frac{2}{3}\left(\frac{q}{q_b}\right)^3\right]
\end{equation}
the numerical prefactor is $\alpha_{\scriptscriptstyle S} = \pi/2$ \cite{larkin1984quantum}. Using these expressions for the barrier height and tunneling action, we
extract relevant phenomenological parameters of the superconductor. With the
experimentally determined value $U/k_{\rm\scriptscriptstyle B} \approx
\SI{2.6}{K}$, we infer a London penetration depth of order $\lambda_{\rm
\scriptscriptstyle L} \approx \SI{2.7}{\upmu m}$. The sheet resistance of the
film is $R^{\scriptscriptstyle\square} \approx \SI{2}{k\Omega}$, and the
coherence length is $\xi \approx \SI{40}{nm}$ \cite{perego2024experimental}.
Combining these values with the expression for $S$, we obtain a tunneling
distance $(3/2) q_b \approx W/8 \approx 3.5 \, \xi$, in good agreement with
the typical barrier thickness extracted from the free-energy landscape in
Eq.~\eqref{eq: Gibbs}.

Figure \ref{Fig1}(d) shows the full temperature dependence of the
dimensionless action $S(T)/\hbar$. For the metastable cubic parabola, this
dependence has been calculated previously \cite{larkin1984quantum} and is
given by
\begin{equation}
 \frac{S(T)}{\hbar} = \frac{S_0}{\hbar}\left[1-\frac{2}{3}
  \left(\frac{T}{T_0}\right)^2\right],
\end{equation}
quadratically decreasing with temperature $T$ to a value $(2/3)S_0/\hbar =
U/k_{\rm\scriptscriptstyle B}T_0$ at the crossover temperature $T_0$.  For temperatures
above $T_0$, the action crosses over to the purely thermal form $S(T \geq
T_0)/\hbar = U/k_{\rm\scriptscriptstyle B} T$. Our experimental results in Fig.~\ref{Fig1}(d) deviate from this prediction in
two respects.  First, the measured action remains approximately constant,
$S(T) \approx S_0$ rather than decreasing by $S_0/3$ as expected upon
approaching $T_0 \approx \SI{100}{mK}$. Second, for a metastable cubic
parabola the dissipative decay is predicted to exhibit a smooth, second-order
type transition at $T_0$ \cite{Affleck1981, larkin1984quantum}. Instead, in
our data, the transition appears markedly sharper.

\begin{figure}[t]
    \centering
    \includegraphics[width=\columnwidth]{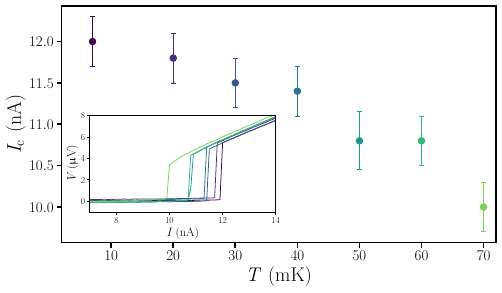}
    \caption{
%
    Critical current $I_c$ at $B=\SI{0}{mT}$ as a function of temperature $T$
    obtained from the sharp $V$--$I$ characteristics shown in the inset;
    the characteristics becomes rounded at higher temperatures $T >
    \SI{70}{mK}$.}
    \label{Fig4}
\end{figure}

These features of $S(T)$ are not universal, but depend strongly on both the
particle dynamics---whether massive or dissipative---and on the shape $V(q)$
of the metastable potential. For example, for a massive particle in a cubic
metastable potential, the action $S(T)$ decreases by only a small fraction,
$1-10\pi/36 \approx 0.13$, up to the (smooth) crossover at $T_0$
\cite{Weiss1999}. Regarding the potential's shape, the curvature of $V(q)$ at
the minimum governs the initial decrease of $S(T)$ with increasing
temperature $T$, whereas the curvature at the barrier top determines the
crossover temperature $T_0$. Moreover, modifications in the overall shape of
$V(q)$ can change the order of the transition, as discussed for the massive
case in Ref.~\cite{Chudnovsky1992phase}.

The free energy landscape $G(y)$ of a Pearl-vortex in a thin film differs
qualitatively from the shape of a cubic parabola.  It rises steeply near the
edges due to vortex--anti-vortex separation arising from image effects and
exhibits a relatively flat maximum, in contrast to the cubic parabola which
has equal curvatures in the minimum and at the barrier maximum.  These
differences lead us to expect a reduced temperature dependence of $S(T)$ and
open the possibility of a first-order like transition.  While quantitative
results are available for massive particle dynamics
\cite{Chudnovsky1992phase}, a corresponding analysis for the dissipative case
remains to be done.  Addressing this problem, together with a more accurate
calculation of the free energy landscape $G(y)$ for vortex entry and exit
beyond the London approximation, constitute interesting problems for
future work.


In conclusion, we have studied vortex dynamics in magic-angle twisted graphene
using a gate-defined Josephson junction as a single-vortex sensor. This
transport-based method enables the detection of individual vortices and their
real-time dynamics. We attribute the observed telegraph-type switching to
vortices traversing the superconducting leads, with thermal activation over
edge barriers giving way to macroscopic quantum tunneling as switching rates
saturate upon decreasing temperature. While such saturation to macroscopic
quantum tunneling is often subject to skepticism---for example, due to
potential failures in temperature equilibration at low temperatures---the
persistence of a clear temperature dependence in the junction critical current
$I_c(T)$ down to our lowest measurement temperature (Fig.~\ref{Fig4}) confirms
that the device is well equilibrated.  Our results provide guidance for the
design and operation of sensitive superconducting devices, such as qubits and
quantum sensors, where vortex fluctuations have a detrimental impact.

\section*{Data availability}
The data supporting the findings of this study, together with the code for
plotting the figures, is available online through the ETH Research Collection
at https://doi.org/10.3929/ethz-c-000794076.

\begin{acknowledgments}
We thank Peter M\"{a}rki and the staff of the ETH cleanroom facility FIRST for
technical support. We acknowledge fruitful discussions with Vladimir Kogan.
Financial support was provided by the European Graphene Flagship Core3
Project, H2020 European Research Council (ERC) Synergy Grant under Grant
Agreement 951541, the European Union’s Horizon 2020 research and innovation
program under grant agreement number 862660/QUANTUM E LEAPS, the European
Innovation Council under grant agreement number 101046231/FantastiCOF, the EU
Cost Action CA21144 (SUPERQUMAP), and NCCR QSIT (Swiss National Science
Foundation, grant number 51NF40-185902).  K.W. and T.T. acknowledge support
from the JSPS KAKENHI (Grant Numbers 21H05233 and 23H02052) and the World
Premier International Research Center Initiative (WPI), MEXT, Japan. C.G.A.
acknowledges support from the Heidi Ras foundation via an ETH Quantum Center
fellowship.  \end{acknowledgments}

\subsection*{Author contributions}
M.P. fabricated the device. T.T. and K.W. supplied the hBN crystals. M.P.,
P.K. and C.G.A. performed the measurements. M.P. and P.K. analysed the data.
V.G. and G.B. developed the theoretical model. M.P. and G.B. wrote the
manuscript, and all authors were involved in the reviewing process. M.P.,
P.K., C.G.A. and A.M.T. discussed the data. M.P., V.G., G.B., K.E. and T.I.
conceived and designed the experiment. T.I. and K.E. supervised the work.

\appendix
\section{Fabrication details}
We fabricated a MAT4G stack using the dry pick-up method~\cite{kim_vdw_2016}.
All the details for the fabrication and tuning of this sample can be found in
\cite{perego2024experimental}.

\section{Measurement setup}
All measurements were carried out in a dilution refrigerator with a base
temperature of $\SI{7}{mK}$. Our measurements are current biased, i.e., we
apply a current and measure the voltage drop in a two-terminal configuration (we correct for contact
resistances). To generate the bias current, we use a home-built d.c.\ source in
series with a $\SI{10}{M\Omega}$ or $\SI{100}{M\Omega}$ resistor.  The
measured voltage is amplified using a home-built low-noise d.c.\ amplifier
(see \cite{marki2017temperature}) and its output is measured with a Hewlett
Packard 3441A digital multimeter. The bottom, top, and finger gates are
connected to home-built low-noise d.c.\ voltage sources.  For the statistical
analysis of vortex fluctuations, the output voltage is further amplified (gain x30k), low-pass filtered at $\SI{1.1}{kHz}$ and recorded in time
using a National Instruments BNC-2110 data acquisition card (DAQ) with a
sampling frequency of $\SI{20}{kHz}$.


\bibliography{Bibliography}

@article{perego2024experimental,
	title = {Experimental detection of vortices in magic-angle graphene},
	author = {Perego, Marta and Galante Agero, Clara and Mestre Tor\'a, Alexandra and Portol\'es, El\'ias and Denisov, Artem O. and Taniguchi, Takashi and Watanabe, Kenji and Gaggioli, Filippo and Geshkenbein, Vadim and Blatter, Gianni and Ihn, Thomas and Ensslin, Klaus}, 
	journal = {Nature Communications},
	year = {2025},
	pages = {10259},
	volume = {16},
	url = {https://doi.org/10.1038/s41467-025-65123-1},
	doi = {10.1038/s41467-025-65123-1}
}

@article{Berezinskii1971, 
       author = {{Berezinski{\v{i}}}, V.~L.},
        title = "{Destruction of Long-range Order in One-dimensional and Two-dimensional Systems having a Continuous Symmetry Group I. Classical Systems}",
      journal = {Soviet Journal of Experimental and Theoretical Physics},
         year = 1971,
        month = jan,
       volume = {32},
        pages = {493},
       adsurl = {https://ui.adsabs.harvard.edu/abs/1971JETP...32..493B}
}

@article{Clem2010,
  title={Josephson junctions in thin and narrow rectangular superconducting strips},
  author={Clem, John R},
  journal={Physical Review B},
  volume={81},
  number={14},
  pages={144515},
  year={2010},
  publisher={APS},
  url = {https://journals.aps.org/prb/abstract/10.1103/PhysRevB.81.144515},
  doi = {10.1103/PhysRevB.81.144515}
}

@article{KT1973,
doi = {10.1088/0022-3719/6/7/010},
url = {https://doi.org/10.1088/0022-3719/6/7/010},
year = {1973},
month = {apr},
publisher = {},
volume = {6},
number = {7},
pages = {1181},
author = {J M Kosterlitz and D J Thouless},
title = {Ordering, metastability and phase transitions in two-dimensional systems},
journal = {Journal of Physics C: Solid State Physics},
}

@article{Rice1965,
  title = {{Superconductivity in One and Two Dimensions}},
  author = {Rice, T. M.},
  journal = {Phys. Rev.},
  volume = {140},
  issue = {6A},
  pages = {A1889--A1891},
  numpages = {0},
  year = {1965},
  month = {Dec},
  publisher = {American Physical Society},
  doi = {10.1103/PhysRev.140.A1889},
  url = {https://link.aps.org/doi/10.1103/PhysRev.140.A1889}
}

@article{Hohenberg1967,
  title = {{Existence of Long-Range Order in One and Two Dimensions}},
  author = {Hohenberg, P. C.},
  journal = {Phys. Rev.},
  volume = {158},
  issue = {2},
  pages = {383--386},
  numpages = {0},
  year = {1967},
  month = {Jun},
  publisher = {American Physical Society},
  doi = {10.1103/PhysRev.158.383},
  url = {https://link.aps.org/doi/10.1103/PhysRev.158.383}
}

@article{MerminWagner1966,
  title = {{Absence of Ferromagnetism or Antiferromagnetism in One- or Two-Dimensional Isotropic Heisenberg Models}},
  author = {Mermin, N. D. and Wagner, H.},
  journal = {Phys. Rev. Lett.},
  volume = {17},
  issue = {22},
  pages = {1133--1136},
  numpages = {0},
  year = {1966},
  month = {Nov},
  publisher = {American Physical Society},
  doi = {10.1103/PhysRevLett.17.1133},
  url = {https://link.aps.org/doi/10.1103/PhysRevLett.17.1133}
}

@article{Perego2026,
  title = {{Vortex Dynamics in Magic-Angle Twisted Graphene}},
  author = {Perego, M. and others},
  journal = {in preparation},
  year = {2026}
}

@article{Devoret1985MQT,
  title = {{Measurements of Macroscopic Quantum Tunneling out of the Zero-Voltage State of a Current-Biased Josephson Junction}},
  author = {Devoret, Michel H. and Martinis, John M. and Clarke, John},
  journal = {Phys. Rev. Lett.},
  volume = {55},
  issue = {18},
  pages = {1908--1911},
  numpages = {0},
  year = {1985},
  month = {Oct},
  publisher = {American Physical Society},
  doi = {10.1103/PhysRevLett.55.1908},
  url = {https://link.aps.org/doi/10.1103/PhysRevLett.55.1908}
}

@article{Pearl1964,
        author = {Pearl, J.},
        doi = {10.1063/1.1754056},
        isbn = {0003-6951},
        journal = {Applied Physics Letters},
        journal1 = {Appl. Phys. Lett.},
        month = {15/08/1964},
        number = {4},
        pages = {65--66},
        title = {{Current distribution in superconducting films carrying quantized fluxoids}},
        url = {https://doi.org/10.1063/1.1754056},
        volume = {5},
        year = {1964},
        year1 = {2004/11/29}}

@article{Abrikosov1957,
        Author = {Abrikosov, A. A.},
        Journal = {{[Zh. Eksp. Teor. Fiz. \textbf{32}, 1442 (1957)]} JETP},
        Pages = {1174},
        Title = {{On the magnetic properties of superconductors of the second group}},
        url = {{http://www.jetp.ac.ru/cgi-bin/dn/e_005_06_1174.pdf}},
        Volume = {5},
        Year = {1957}}

@article{BeanLivingston1964,
        author = {Bean, C. P. and Livingston, J. D.},
        date = {1964/01/06/},
        day = {06},
        doi = {10.1103/PhysRevLett.12.14},
        id = {10.1103/PhysRevLett.12.14},
        j1 = {PRL},
        journal = {Physical Review Letters},
        journal1 = {Phys. Rev. Lett.},
        month = {01},
        number = {1},
        pages = {14--16},
        publisher = {American Physical Society},
        title = {{Surface Barrier in Type-II Superconductors}},
        url = {https://link.aps.org/doi/10.1103/PhysRevLett.12.14},
        volume = {12},
        year = {1964}}

@article{Affleck1981,
  title = {{Quantum-Statistical Metastability}},
  author = {Affleck, Ian},
  journal = {Phys. Rev. Lett.},
  volume = {46},
  issue = {6},
  pages = {388--391},
  numpages = {0},
  year = {1981},
  month = {Feb},
  publisher = {American Physical Society},
  doi = {10.1103/PhysRevLett.46.388},
  url = {https://link.aps.org/doi/10.1103/PhysRevLett.46.388}
}

@article{Clem2011,
  title = {{Effect of nearby Pearl vortices upon the ${I}_{c}$ versus $B$ characteristics of planar Josephson junctions in thin and narrow superconducting strips}},
  author = {Clem, John R.},
  journal = {Phys. Rev. B},
  volume = {84},
  issue = {13},
  pages = {134502},
  numpages = {7},
  year = {2011},
  month = {Oct},
  publisher = {American Physical Society},
  doi = {10.1103/PhysRevB.84.134502},
  url = {https://link.aps.org/doi/10.1103/PhysRevB.84.134502}
}

@article{KoganMints2014,
  title={{Interaction of Josephson junction and distant vortex in narrow thin-film superconducting strips}},
  author={Kogan, V.G. and Mints, R.G.},
  journal={Physical Review B},
  volume={89},
  pages={014516},
  year={2014},
  publisher={APS},
  doi = {10.1103/PhysRevB.89.014516},
  url = {https://journals.aps.org/prb/pdf/10.1103/PhysRevB.89.014516}
}

@article{KoganMints_PC2014,
  title={{Manipulating Josephson junctions in thin-films by nearby vortices}},
  author={Kogan, V.G. and Mints, R.G.},
  journal={Physica C},
  volume={502},
  pages={58},
  year={2014},
  publisher={Elsevier},
  doi = {10.1016/j.physc.2014.04.039},
  url = {https://www.sciencedirect.com/science/article/pii/S0921453414001476?via%3Dihub}
}

@article{Freedman_Diaconis_1981,
	author = {Freedman, David and Diaconis, Persi},
	date = {1981/12/01},
	doi = {10.1007/BF01025868},
	id = {Freedman1981},
	isbn = {1432-2064},
	journal = {Zeitschrift f{\"u}r Wahrscheinlichkeitstheorie und Verwandte Gebiete},
	pages = {453--476},
	title = {On the histogram as a density estimator : {{$L_2$}} theory},
	url = {https://doi.org/10.1007/BF01025868},
	volume = {57},
    month = {Dec},
	year = {1981}}

@article{Yuzhelevski_2000,
    author = {Yuzhelevski, Y. and Yuzhelevski, M. and Jung, G.},    title = {Random telegraph noise analysis in time domain},    journal = {Review of Scientific Instruments},    volume = {71},        pages = {1681-1688},    year = {2000},    month = {Apr},    issn = {0034-6748},
    doi = {10.1063/1.1150519},    url = {https://doi.org/10.1063/1.1150519}
}

@article{Malozemoff1990,
  title = {{Universality in the current decay and flux creep of Y-Ba-Cu-O high-temperature superconductors}},
  author = {Malozemoff, Alexis P. and Fisher, Matthew P. A.},
  journal = {Phys. Rev. B},
  volume = {42},
  issue = {10},
  pages = {6784--6786},
  numpages = {0},
  year = {1990},
  month = {Oct},
  publisher = {American Physical Society},
  doi = {10.1103/PhysRevB.42.6784},
  url = {https://link.aps.org/doi/10.1103/PhysRevB.42.6784}
}

@article{Koshelev1990, 
  title = {{The role of surface effects in magnetization of high-Tc superconductors}},
  author = {Kopylov, V.N. and others},
  journal = {Physica C}, 
  volume = {170},
  pages = {291},
  year = {1990},
  publisher = {Elsevier}, 
  doi = {10.1016/0921-4534(90)90326-A},
  url = {https://doi.org/10.1016/0921-4534(90)90326-A}
}

@article{Naaman_2006,
  title = {{Poisson Transition Rates from Time-Domain Measurements with a Finite Bandwidth}},  author = {Naaman, O. and Aumentado, J.},  journal = {Phys. Rev. Lett.},  volume = {96},  issue = {10},  year = {2006},  month = {Mar},  publisher = {American Physical Society},  doi = {10.1103/PhysRevLett.96.100201},  url = {https://link.aps.org/doi/10.1103/PhysRevLett.96.100201}
}

@article{Stejic_1994,   
title = {Effect of geometry on the critical currents of thin films},  
author = {Stejic, G. and Gurevich, A. and Kadyrov, E. and Christen, D. and Joynt, R. and Larbalestier, D. C.},   
journal = {Phys. Rev. B},   
volume = {49},   
issue = {2},   
pages = {1274--1288},   
numpages = {0},   
year = {1994},   
month = {Jan},   
publisher = {American Physical Society},   
doi = {10.1103/PhysRevB.49.1274},    
url = {https://link.aps.org/doi/10.1103/PhysRevB.49.1274}
}

@article{Chudnovsky1992phase,
  title={Phase transitions in the problem of the decay of a metastable state},
  author={Chudnovsky, Eugene M},
  journal={Physical Review A},
  volume={46},
  number={12},
  pages={8011},
  year={1992},
  publisher={APS}
}

@article{glazman1984,
  title={Possibility of quantum tunneling of vortices in thin superconducting films},
  author={Glazman, LI and Fogel’, N Ya},
  journal={Soviet Journal of Low Temperature Physics},
  volume={10},
  number={1},
  pages={51--52},
  year={1984},
  publisher={American Institute of Physics}
}

@article{caldeira1983quantum,
  title={Quantum tunnelling in a dissipative system},
  author={Caldeira, Amir O and Leggett, Anthony J},
  journal={Annals of physics},
  volume={149},
  number={2},
  pages={374--456},
  year={1983},
  publisher={Academic Press}
}

@article{larkin1984quantum,
  title={Quantum-mechanical tunneling sith dissipation. the pre-exponential factor},
  author={Larkin, A and Ovchinnikov, Yu N},
  journal={Zh. Eksp. Teor. Fiz},
  volume={86},
  pages={726},
  year={1984}
}

@article{wu2019phonon,
  title={{Phonon-induced giant linear-in-T resistivity in magic angle twisted bilayer graphene: Ordinary strangeness and exotic superconductivity}},
  author={Wu, Fengcheng and Hwang, Euyheon and Das Sarma, Sankar},
  journal={Physical Review B},
  volume={99},
  number={16},
  pages={165112},
  year={2019},
  publisher={APS}
}

@article{oh2021evidence,
  title={Evidence for unconventional superconductivity in twisted bilayer graphene},
  author={Oh, Myungchul and Nuckolls, Kevin P and Wong, Dillon and Lee, Ryan L and Liu, Xiaomeng and Watanabe, Kenji and Taniguchi, Takashi and Yazdani, Ali},
  journal={Nature},
  volume={600},
  number={7888},
  pages={240--245},
  year={2021},
  publisher={Nature Publishing Group UK London}
}

@article{cao2018correlated,
  title={{Correlated insulator behaviour at half-filling in magic-angle graphene superlattices}},
  author={Cao, Yuan and Fatemi, Valla and Demir, Ahmet and Fang, Shiang and Tomarken, Spencer L and Luo, Jason Y and Sanchez-Yamagishi, Javier D and Watanabe, Kenji and Taniguchi, Takashi and Kaxiras, Efthimios and others},
  journal={Nature},
  volume={556},
  number={7699},
  pages={80--84},
  year={2018},
  publisher={Nature Publishing Group UK London},
  doi={10.1038/nature26154}
}

@article{lu2019superconductors,
  title={{Superconductors, orbital magnets and correlated states in magic-angle bilayer graphene}},
  author={Lu, Xiaobo and Stepanov, Petr and Yang, Wei and Xie, Ming and Aamir, Mohammed Ali and Das, Ipsita and Urgell, Carles and Watanabe, Kenji and Taniguchi, Takashi and Zhang, Guangyu and others},
  journal={Nature},
  volume={574},
  number={7780},
  pages={653--657},
  year={2019},
  publisher={Nature Publishing Group UK London},
  doi={10.1038/s41586-019-1695-0}
}

@article{hao2021electric,
  title={{Electric field--tunable superconductivity in alternating-twist magic-angle trilayer graphene}},
  author={Hao, Zeyu and Zimmerman, AM and Ledwith, Patrick and Khalaf, Eslam and Najafabadi, Danial Haie and Watanabe, Kenji and Taniguchi, Takashi and Vishwanath, Ashvin and Kim, Philip},
  journal={Science},
  volume={371},
  number={6534},
  pages={1133--1138},
  year={2021},
  publisher={American Association for the Advancement of Science},
  doi={10.1126/science.abg0399}
}

@article{park2021tunable,
  title={{Tunable strongly coupled superconductivity in magic-angle twisted trilayer graphene}},
  author={Park, Jeong Min and others},
  journal={Nature},
  volume={590},
  number={7845},
  pages={249--255},
  year={2021},
  publisher={Nature Publishing Group UK London},
  doi={10.1038/s41586-021-03192-0}
}

@article{park2022robust,
  title={{Robust superconductivity in magic-angle multilayer graphene family}},
  author={Park, Jeong Min and others},
  journal={Nature Materials},
  volume={21},
  number={8},
  pages={877--883},
  year={2022},
  publisher={Nature Publishing Group UK London},
  doi={10.1038/s41563-022-01287-1}
}

@article{burg2022emergence,
  title={{Emergence of correlations in alternating twist quadrilayer graphene}},
  author={Burg, G William and Khalaf, Eslam and Wang, Yimeng and Watanabe, Kenji and Taniguchi, Takashi and Tutuc, Emanuel},
  journal={Nature Materials},
  volume={21},
  number={8},
  pages={884--889},
  year={2022},
  publisher={Nature Publishing Group UK London},
  doi={10.1038/s41563-022-01286-2}
}

@article{zhang2022promotion,
  title={{Promotion of superconductivity in magic-angle graphene multilayers}},
  author={Zhang, Yiran and Polski, Robert and Lewandowski, Cyprian and Thomson, Alex and Peng, Yang and Choi, Youngjoon and Kim, Hyunjin and Watanabe, Kenji and Taniguchi, Takashi and Alicea, Jason and others},
  journal={Science},
  volume={377},
  number={6614},
  pages={1538--1543},
  year={2022},
  publisher={American Association for the Advancement of Science},
  doi={10.1126/science.abn8585}
}

@article{diez2023symmetry,
  title={{Symmetry-broken Josephson junctions and superconducting diodes in magic-angle twisted bilayer graphene}},
  author={D{\'\i}ez-M{\'e}rida, J and D{\'\i}ez-Carl{\'o}n, A and Yang, SY and Xie, Y-M and Gao, X-J and Senior, J and Watanabe, K and Taniguchi, T and Lu, X and Higginbotham, Andrew P and others},
  journal={Nature Communications},
  volume={14},
  number={1},
  pages={2396},
  year={2023},
  publisher={Nature Publishing Group UK London},
  doi={10.1038/s41467-023-38005-7}
}

@article{portoles2022tunable,
  title={{A tunable monolithic SQUID in twisted bilayer graphene}},
  author={Portol{\'e}s, El{\'\i}as and Iwakiri, Shuichi and Zheng, Giulia and Rickhaus, Peter and Taniguchi, Takashi and Watanabe, Kenji and Ihn, Thomas and Ensslin, Klaus and de Vries, Folkert K},
  journal={Nature Nanotechnology},
  volume={17},
  number={11},
  pages={1159--1164},
  year={2022},
  publisher={Nature Publishing Group UK London},
  doi={10.1038/s41565-022-01222-0}
}

@article{wakiri2024tunable,
  title={{Tunable quantum interferometer for correlated moir{\'e} electrons}},
  author={Iwakiri, Shuichi and Mestre-Tor{\`a}, Alexandra and Portol{\'e}s, El{\'\i}as and Visscher, Marieke and Perego, Marta and Zheng, Giulia and Taniguchi, Takashi and Watanabe, Kenji and Sigrist, Manfred and Ihn, Thomas and others},
  journal={Nature Communications},
  volume={15},
  number={1},
  pages={390},
  year={2024},
  publisher={Nature Publishing Group UK London},
  doi={10.1038/s41467-023-44671-4}
}

@article{cao2018unconventional,
  title={{Unconventional superconductivity in magic-angle graphene superlattices}},
  author={Cao, Yuan and Fatemi, Valla and Fang, Shiang and Watanabe, Kenji and Taniguchi, Takashi and Kaxiras, Efthimios and Jarillo-Herrero, Pablo},
  journal={Nature},
  volume={556},
  number={7699},
  pages={43--50},
  year={2018},
  publisher={Nature Publishing Group},
  url={https://www.nature.com/articles/nature26160},
  doi= {10.1038/nature26160}
}

@Article{deVries2021,
author={de Vries, Folkert K.
and Portol{\'e}s, El{\'i}as
and Zheng, Giulia
and Taniguchi, Takashi
and Watanabe, Kenji
and Ihn, Thomas
and Ensslin, Klaus
and Rickhaus, Peter},
title={{Gate-defined Josephson junctions in magic-angle twisted bilayer graphene}},
journal={Nature Nanotechnology},
year={2021},
month={Jul},
day={01},
volume={16},
number={7},
pages={760-763},
issn={1748-3395},
doi={10.1038/s41565-021-00896-2}
}

@article{diez2025probing,
  title = {{Probing the Flat-Band Limit of the Superconducting Proximity Effect in Twisted Bilayer Graphene Josephson Junctions}},
  author = {D\'{\i}ez-Carl\'on, A. and D\'{\i}ez-M\'erida, J. and Rout, P. and Sedov, D. and Virtanen, P. and Banerjee, S. and Penttil\"a, R. P. S. and Altpeter, P. and Watanabe, K. and Taniguchi, T. and Yang, S.-Y. and Law, K. T. and Heikkil\"a, T. T. and T\"orm\"a, P. and Scheurer, M. S. and Efetov, D. K.},
  journal = {Phys. Rev. X},
  volume = {15},
  issue = {4},
  pages = {041033},
  numpages = {11},
  year = {2025},
  month = {Nov},
  publisher = {American Physical Society},
  doi = {10.1103/ccb4-tqxq},
  url = {https://link.aps.org/doi/10.1103/ccb4-tqxq}
}

@article{rothstein2025gate,
  title={{Gate-Defined Single-Electron Transistors in Twisted Bilayer Graphene}},
  author={Rothstein, Alexander and Fischer, Ammon and Achtermann, Anthony and Icking, Eike and Hecker, Katrin and Banszerus, Luca and Otto, Martin and Trellenkamp, Stefan and Lentz, Florian and Watanabe, Kenji and others},
  journal={Nano Letters},
  volume={25},
  number={16},
  pages={6429--6437},
  year={2025},
  publisher={ACS Publications},
  doi={10.1021/acs.nanolett.4c06492}
}

@Article{Rodan-Legrain2021,
author={Rodan-Legrain, Daniel
and Cao, Yuan
and Park, Jeong Min
and de la Barrera, Sergio C.
and Randeria, Mallika T.
and Watanabe, Kenji
and Taniguchi, Takashi
and Jarillo-Herrero, Pablo},
title={{Highly tunable junctions and non-local Josephson effect in magic-angle graphene tunnelling devices}},
journal={Nature Nanotechnology},
year={2021},
month={Jul},
day={01},
volume={16},
number={7},
pages={769-775},
issn={1748-3395},
doi={10.1038/s41565-021-00894-4}
}

@article{zheng2024gate,
  title={{Gate-defined superconducting channel in magic-angle twisted bilayer graphene}},
  author={Zheng, Giulia and Portol{\'e}s, El{\'\i}as and Mestre-Tor{\`a}, Alexandra and Perego, Marta and Taniguchi, Takashi and Watanabe, Kenji and Rickhaus, Peter and de Vries, Folkert K and Ihn, Thomas and Ensslin, Klaus and others},
  journal={Physical Review Research},
  volume={6},
  number={1},
  pages={L012051},
  year={2024},
  publisher={APS},
  doi={10.1103/PhysRevResearch.6.L012051}
}

@article{Ronen2025,
  title = {{Competing Orbital Magnetism and Superconductivity in electrostatically defined Josephson Junctions of Alternating Twisted Trilayer Graphene}},
  author = {Ronen, Y. and Bhardwaj, V. and Rajagopal, L. and Arici, L. and Bocarsly, M. and Ilin, A. and Shavit, G. and Watanabe, K. and Taniguchi, T. and Oreg, Y. and Holder, T.},
  journal = {unpublished},
  volume = {},
  pages = {},
  year = {2025},
  publisher = {Elsevier},
  doi = {10.21203/rs.3.rs-5756127/v1}, 
  url = {https://doi.org/10.21203/rs.3.rs-5756127/v1}
}

@article{Kim2024,
  title={{Superfluid stiffness of twisted trilayer graphene superconductors}},
  author={Banerjee, Abhishek and Hao, Zeyu and Kreidel, Mary and Ledwith, Patrick and Phinney, Isabelle and Park, Jeong Min and Zimmerman, Andrew and Wesson, Marie E and Watanabe, Kenji and Taniguchi, Takashi and others},
  journal={Nature},
  volume={638},
  number={8049},
  pages={93--98},
  year={2025},
  publisher={Nature Publishing Group UK London},
  doi={10.1038/s41586-024-08444-3}
}

@article{Oliver2024,
  title={{Superfluid stiffness of magic-angle twisted bilayer graphene}},
  author={Tanaka, Miuko and Wang, Joel {\^I}-j and Dinh, Thao H and Rodan-Legrain, Daniel and Zaman, Sameia and Hays, Max and Almanakly, Aziza and Kannan, Bharath and Kim, David K and Niedzielski, Bethany M and others},
  journal={Nature},
  volume={638},
  number={8049},
  pages={99--105},
  year={2025},
  publisher={Nature Publishing Group UK London},
  doi={10.1038/s41586-024-08494-7}
}

@article{portoles2024quasiparticle,
  title={{Quasiparticle and superfluid dynamics in Magic-Angle Graphene}},
  author={Portol{\'e}s, El{\'\i}as and Perego, Marta and Volkov, Pavel A and Toschini, Mathilde and Kemna, Yana and Mestre-Tor{\`a}, Alexandra and Zheng, Giulia and Denisov, Artem O and Vries, Folkert K de and Rickhaus, Peter and others},
  journal={Nature Communications},
  volume={16},
  number={1},
  pages={1--9},
  year={2025},
  publisher={Nature Publishing Group},
doi = {10.1038/s41467-025-58325-0}
}

@book{Weiss1999,
  title={{Quantum Dissipative Systems (2nd edition)}},
  author={Ulrich Weiss},
  year={1999},
  publisher={World Scientific Publishing},
  address={Singapore}
}

@book{Tinkham2004,
  title={{Introduction to Superconductivity}},
  author={Tinkham, Michael},
  year={2004},
  publisher={Dover Publications},
  address={Mineola, NY}
}

@article{bardeen1965theory,
  title={Theory of the motion of vortices in superconductors},
  author={Bardeen, John and Stephen, MJ},
  journal={Physical Review},
  volume={140},
  number={4A},
  pages={A1197},
  year={1965},
  publisher={APS}
}

@article{jha2025large,
  title={Large tunable kinetic inductance in a twisted graphene superconductor},
  author={Jha, Rounak and Endres, Martin and Watanabe, Kenji and Taniguchi, Takashi and Banerjee, Mitali and Sch{\"o}nenberger, Christian and Karnatak, Paritosh},
  journal={Physical Review Letters},
  volume={134},
  number={21},
  pages={216001},
  year={2025},
  publisher={APS}
}

@article{marki2017temperature,
  title={{Temperature-stabilized differential amplifier for low-noise DC measurements}},
  author={M{\"a}rki, Peter and Braem, Beat A and Ihn, T},
  journal={Review of Scientific Instruments},
  volume={88},
  number={8},
  year={2017},
  publisher={AIP Publishing},
  doi={10.1063/1.4997963}
}

@article{kim_vdw_2016,
    author={Kim, Kyounghwan and Yankowitz, Matthew and Fallahazad, Babak and Kang, Sangwoo and Movva, Hema C. P. and Huang, Shengqiang and Larentis, Stefano and Corbet, Chris M. and Taniguchi, Takashi and Watanabe, Kenji and Banerjee, Sanjay K. and LeRoy, Brian J. and Tutuc, Emanuel},
    title={{van der Waals Heterostructures with High Accuracy Rotational Alignment}},
    journal={Nano Letters},
    year={2016},
    publisher={American Chemical Society},
    volume={16},
    number={3},
    pages={1989-1995},
    issn={1530-6984},
   doi={10.1021/acs.nanolett.5b05263}
}

\end{document}